\documentclass[twocolumn,pra,aps,superscriptaddress,showpacs,amsmath,amssymb,floatfix]{revtex4-1}
\usepackage{graphicx}
\usepackage{amssymb}
\usepackage{amsmath}
\usepackage{epsfig}
\usepackage{color}
\usepackage{mathtools}
\usepackage[colorlinks,linkcolor=blue,anchorcolor=blue,c itecolor=blue,urlcolor=blue]{hyperref}
\usepackage{physics}
\usepackage{ulem}

\begin{document}

\title{Characterizing Topological Phase Transition  in  Non-Hermitian Systems}

\author{ZhaoXiang Fang}

\affiliation{School of Physical Science and Technology, Xinjiang University, Urumqi, 830046, China}
\affiliation{School of physics, University of Science and Technology of China, Hefei, 230026, China}

\author{Yongxu Fu}
\email{yongxufu@zjnu.edu.cn}
\affiliation{Department of Physics, Zhejiang Normal University, Jinhua 321004, China}

\author{Guang-Can Guo}
\affiliation{School of physics, University of Science and Technology of China, Hefei, 230026, China}
\affiliation{Key Lab of Quantum Information, Chinese Academy of Sciences, School of physics, University of Science and Technology of China, Hefei, 230026, China}

\author{Long Xiong}
\email{XiongLong_phy@outlook.com}
\affiliation{International Center for Quantum Materials, School of Physics, Peking University, Beijing, 100871, China}

\date{\today }

\begin{abstract}
We propose and present a concept of Topological Distance (TD), obtained from the integration of trace distance over the generalized Brillouin zone, in order to characterize the topological transitions of non-Hermitian systems. Specifically, such a quantity is used to measure the overall dissimilarity between eigen wavefunctions upon traversing all possible matter states, and confirms the phase boundaries through observing the divergences of both TD and its partial derivatives; we clarify its origin and also offer a theoretical explanation. The method is developed to characterize the non-Hermitian topology in a novel way, and shows its generality and effectiveness in  1D non-Hermitian Kitaev system, non-Hermitian Hamiltonians under periodic  or
open boundary conditions, and even generalizable to higher-order topological systems, providing a novel perspective to understand topological physics.  
\end{abstract}

\maketitle

Topology describes a kind of geometry that objects can deform into each other, and can also be used to characterize phases of matter in physics, including material with the quantum Hall effect \cite{von1980, tknn, haldane1988}, topological insulators and topological semimetals \cite{PhysRevLett.95.146802, PhysRevLett.95.226801}. These systems are characterized by a variety of topological invariants that dictate the presence of boundary states via the bulk-boundary correspondence. For instance, Chern number, derived in a Berry phase that integrated curvature over Brillouin zone, could classify the features of two dimensional (2D) topological systems  \cite{hatsugai1993chern,kane2005z}; winding number could be formulated to describe the topological properties for one-dimensional (1D) Su-Schrieffer-Heeger (SSH) model \cite{su1980soliton}; the $Z_2$ topological index was defined for time reversal invariant Hamiltonians that supported the transport of spin in gapless edge states \cite{schnyder2009classification}. The topological invariants were so significant that one could better understand many body phases having bulk energy gaps \cite{qi2011topological,schnyder2008classification}.

Studies of topological invariants mainly emphasize on the Hermitian Hamiltonians. In recent decade, a growing interest to investigate the basic topology and dynamics under the non-Hermitian Hamiltonians has been induced \cite{bergholtz2021exceptional,gong2018topological,kawabata2019symmetry,shen2018topological,yao2018edge,song2019non,lee2016anomalous,kunst2018,yao201802,yokomizo2019,origin2020,slager2020,yang2020,zhang2020,guo2021exact,edgeburst2022,kawabata2019second,lee2019ho,fu2021,st2022}, such as non-Hermitian skin effect under open boundary condition (OBC) and periodic boundary condition (PBC). Typically, non-Hermitian Hamiltonians possess complex eigenvalues, scale-free localization \cite{fu2023hybrid,molignini2023anomalous}, and exceptional points (EPs) occur when two or more eigenmodes coalesce \cite{kawabata2019,jones2020,xue2020dirac,yang2021,denner2021,fu2022,mandal2021ep,delplace2021ep,liu2021ep}. However, it is challenging to generalize the traditional invariants to non-Hermitian Hamlitonians, as the complex energy spectra makes it difficult to define band gaps clearly and skin effect suggests a "generalized bulk-boundary correspondence" instead of conventional one. Though various topological invariants, such as non-Bloch winding number (Chern number) defined via complex wave vectors in 1D (2D) and vorticity obtained from the winding of the complex energy spectrum around EPs, have been introduced in recent works \cite{shen2018topological,yao201802,yao2018edge,yokomizo2019,lee2016anomalous}, they are either confined to non-Hermitian Hamlitonians of certain models  or restricted by specific symmetries (e.g. PT or chiral symmetry), and some of their analytic expression requires a more reduced form when determining the boundary states.  Thus, it remains an open question: is there a new quantity or method that characterizes the non-Hermitian topology in a more generalized and effective way?

Generally, traditional quantum phase transition (QPT) features with localized discontinuities in ground state \cite{jozsa1994fidelity,nandi2018two,liang2019quantum,gu2010fidelity}; whereas, the topological phase transition (TPT) should be determined through capturing the global properties. Though trace distance quantifies the distance between wave functions in characterizing QPT, it is not affected by non-Hermiticity \cite{jozsa1994fidelity,li2012superfidelity,rastegin2007trace,zhang2019subsystem,brito2018quantifying}; inspired by the Chern number, if it concerns integration over the generalized Brillouin zone (GBZ) and measures the overall dissimilarity between wavefunctions in momentum space, such quantity can be naturally introduced and redefined in characterizing TPT of non-Hermitian topology. Notably, this is an intriguing thought and has not been explored before.

In this letter, we propose and present the concept of “Topological Distance (TD)”, which is derived through intergrating trace distance over the entire generalized Brillouin zone, enabling the characterization of phase transitions in non-Hermitian topology.  The TD is employed to calculate the general difference between wavefunctions that traverse all possible object states of matter, and determines the transition points through studying the divergences of both TD and its partial derivatives, distinguish from the conventional topological invariants. We clarify the origin and then give a theoretical explanation. The method based on TD is applicable to non-Hermitian topological systems independent of symmetry, where their GBZ are well defined, including 1D PT-symmetric SSH under PBC or OBC, and 2D or 3D second-order topology in the presence of non-Hermiticity. Such a method is developed to characterize the topological transitions
of non-Hermitian system in a generalized and effective manner, providing a novel insight to investigate the topological physics.
~\\

{\it Topological Distance.-}The concept of TD is mainly based on trace distance, which is used to characterize the QPT through observing the existence of a localized discontinuity.  The trace distance between two mixed states is defined as:

\begin{align}
	D\left(\rho, \sigma \right)=\frac{1}{2}\left\|\rho-\sigma \right\|_1
\end{align}
where D denotes the trace distance; $\left\| \cdot \right\|_1$ represents the Schatten-1 norm; $\rho$ and $\sigma$ refer to the density matrices for the mixed states. The relation between trace distance and fidelity can be derived:
\begin{align}
	1-F\left(\rho, \sigma \right) \leq D\left(\rho, \sigma \right) \leq \sqrt{1-F^2\left(\rho, \sigma \right)},
\end{align}

Note that the upper bound inequality becomes an equality when $\rho$ and $\sigma$  are pure states\cite{jozsa1994fidelity}. While for pure states, we employ manifolds that denote two eigenstates geometrically in Fig. 1(a), i.e., $\phi$ and $\psi$; the two are topologically equivalent when in identical states.  Here, inspired by Chern number, TD is proposed to characterize TPT and describes as:

\begin{align}
	D=\sqrt{1-|\bra*{\phi}\ket*{\psi}|^2},
\end{align}

\begin{align}
	T_d=\int_{GBZ} D dk;   \quad T_d^{(n)} = \int_{GBZ} \frac{\partial^n D}{\partial O^n}  dk; 
	\label{eq7}
\end{align}

where  $T_d$ refers to TD for simplicity of calculation; $k$ denotes the parameter in momentum space; the $T_d^{(n)}$ can be obtained by taking the partial derivatives with respect to $O$, which represents the parameter that governs the phase boundary. $\phi$ denoted as a “reference” wavefunction, is set in a random state initially and keeps unchanged during calculation; $\psi$ presents as an “object” state that traverses a targeted region with a constant range.

Theoretically, similar to the definition of Chern number, the $T_d$ is obtained by the integration of $D$ over entire GBZ, while the latter quantifies the general dissimilarity between wavefunctions accumulated among all possible matter states in a given region, thus confirming the presence of phase transitions explicitly, instead of topological index. Moreover, the $n$-th partial derivative $T_d^{(n)}$ is employed to describe divergent behaviors further in the vicinity of phase boundary; normally, the $T_d^{\prime \prime}$($n$=2) is capable enough to show the divergences.

\begin{figure}
	\centering
	\includegraphics[width=0.48\textwidth]{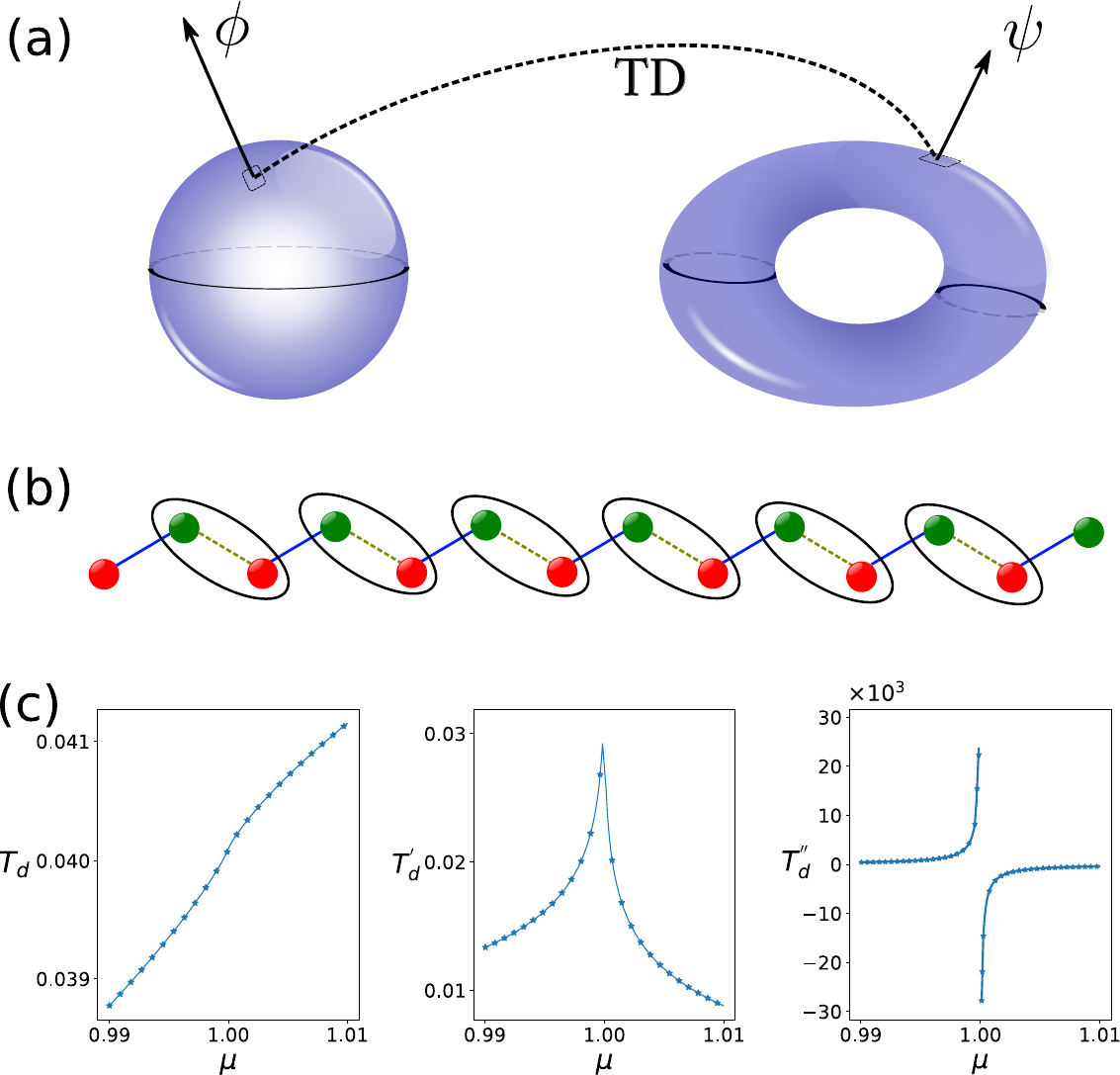}
	\caption{The mapping function between two matter states and the validity of TD verified in 1D non-Hermitian
Kitaev model. (a) the manifolds  denote $\phi$ and $\psi$, and the general dissimilarity between states are mapped  by $T_d$. (b) schematic diagram of 1D non-Hermitian Kitaev model. (c) $T_d$ with its partial derivatives are calculated for non-Hermitian
Kitaev model and $\mu$ refers to the chemical potential.}
	\label{fig1}
\end{figure}

~\\

{\it Generalized to the non-Hermitian Hamiltonians.-}As the complex eigenvalues and skin effect of non-Hermitian system disrupt the conventional bulk-boundary correspondence, traditional invariants cannot correctly determine the phase boundary between topological and trivial states\cite{bergholtz2021exceptional,gong2018topological,shen2018topological,yao2018edge,song2019non,lieu2018topological}. However, $T_d$ is not influenced by the non-Hermiticity in our work and the imaginary part of complex-valued wavefunction shows no effect on divergent property during calculation, thus such quantity can be extended to characterize the topological phase transition in non-Hermitian systems,  according to the definition of $T_d$ and its partial derivatives in Eq. 4.

For non-Hermiticity is introduced by incorporating NH hopping terms or gain/loss terms into Hermitian Hamiltonians \cite{liu2020gain,kawabata2019symmetry}, one consider the eigenvalue equations of non-Hermitian Hamiltonian as 
\begin{align}
	H_k \ket{\varphi_n} = E_n \ket{\varphi_n},\quad H_k^{\dagger} \ket{\phi_n} = E_{n}^{*} \ket{\phi_n}.
\end{align}
where we have two forms of eigen-equations and  Hamiltonians. 
~\\

We start our discussions with 1D non-Hermitian Kitaev model \cite{kawabata2019symmetry}, as shown schematically in Fig. 1(b), which can be described by the Hamiltonian: 
\begin{align}
	H_k=\left(\begin{array}{cc}
		-2t\cos (k)-\mu+i \gamma & \alpha \sin (k) \\
		\alpha \sin (k) & 2t\cos (k)+\mu-i \gamma, \\
	\end{array}\right)
\end{align}
where $t$ represents the nearest-neighbor hopping amplitude, $\mu$ denotes the chemical potential controlling the filling level, $\alpha$ stands for the $p$-wave superconducting pairing strength between neighboring sites, $k$ refers to the momentum in the Brillouin zone, and the parameter $\gamma$ introduces non-Hermitian strength, modeling dissipation or amplification. In this model, $\mu$ is equivalent to O in the Eq. 4. The transition points describe as:
\begin{align}
	\mu=\pm 2t \sqrt{1-\frac{\gamma^2}{\alpha^2}}, \quad \gamma^2 \le \alpha^2
    \label{p_wave_phase}
\end{align}

As demonstrated in Fig. 1(c), we observe the dynamics of $T_d^{(n)}$   as a function of $\mu$, and confirm a critical point locating at $\mu=1.00$, when the parameters are set to be $t=0.5$, $\alpha=0.8$ and $\gamma=0.01$, agreeing well with Ref\cite{kawabata2019symmetry}.  In the following, to further validate the method, we will perform numerical calculations of $T_d$ and its derivatives in a variety of non-Hermitian topological systems.

\begin{figure}
	\centering
	\includegraphics[width=0.48\textwidth]{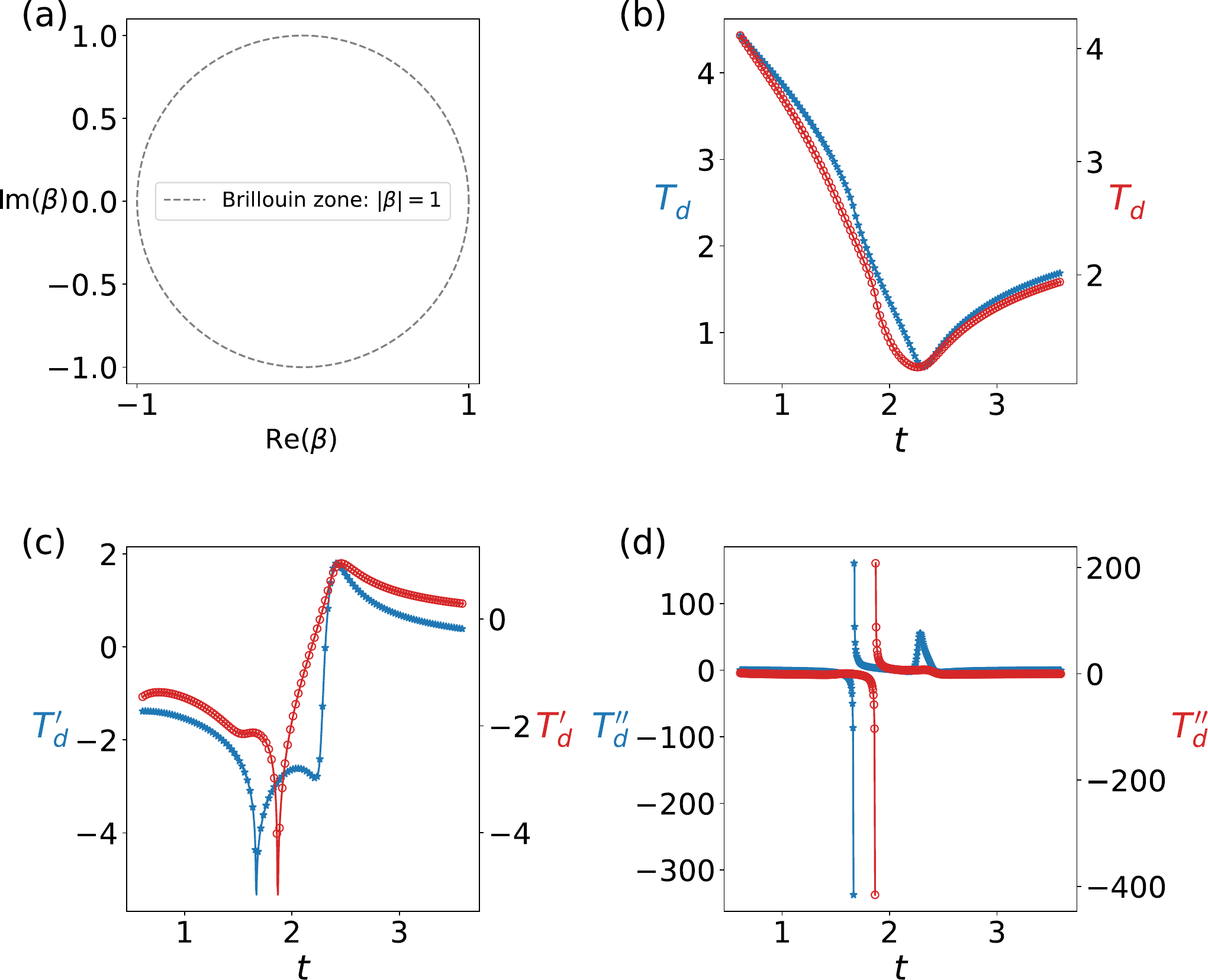}
	\caption{$T_d$ and its partial deriveatives for NH Hamiltonians  under PBCs. (a) depicts the  Brillouin zone; (b)-(d) denote the $T_{d}$, its first and second-order derivative respectively; red and blue line signify the system with or without next-nearest-neighbor interaction. }
	\label{fig2}
\end{figure}

\begin{figure}
	\centering
	\includegraphics[width=0.48\textwidth]{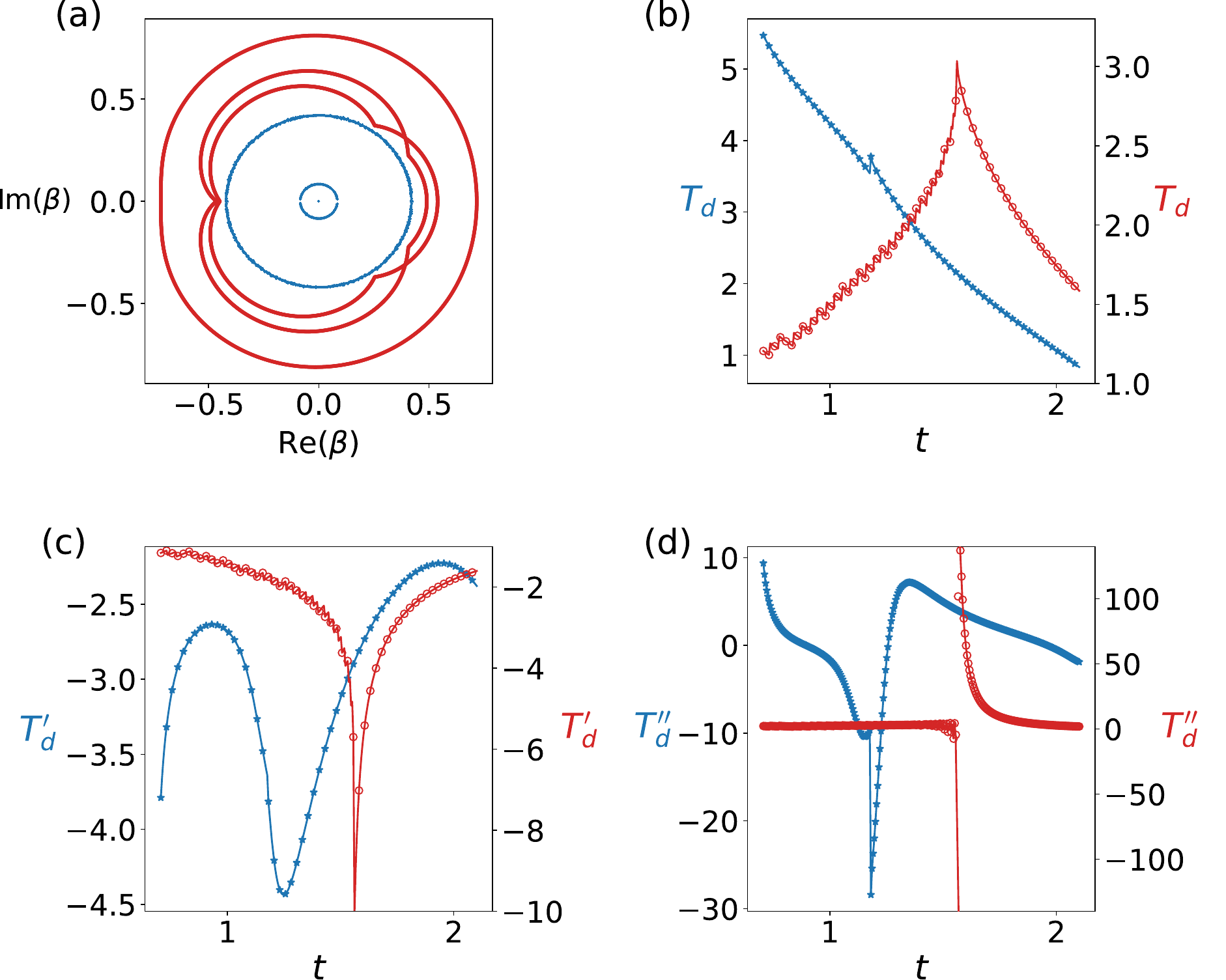}
	\caption{$T_d$ and its partial derivatives for NH Hamiltonians under OBCs. (a) shows its generalized Brillouin zone; (b)-(d) denote the $T_d$, $T_d^{\prime}$ and $T_d^{\prime \prime}$ of the NH systems respectively; red and blue line signify the system with or without next-nearest-neighbor interaction.}
	\label{fig3}
\end{figure}

~\\

{\it Periodic Boundary Conditions.-}We first study the divergent property of non-Hermitian Hamiltonian under PBCs, and  the method is employed  to verify its validity. Here, a 1D PT-symmetric NH Su-Schrieffer-Heeger (SSH) model is adopted and its Hamiltonian reads as \cite{yao2018edge}.
\begin{align}
	H_k=\left(\begin{array}{cc}
		0 & \beta(k)\\
		\beta^{*}(k) & 0 \\
	\end{array}\right),
	\label{pt-nh}
\end{align}
with
\begin{align}
	& \beta(k) = \frac{\gamma}{2} + t + (t^{'} +t_3) \cos(k) - i (t^{'}-t_3)  \sin(k), \\
	& \beta^{*}(k)= -\frac{\gamma}{2} + t + (t^{'} +t_3) \cos(k) + i (t^{'}-t_3) \sin(k) \nonumber
	\label{pt-H}
\end{align}
where  $t$ and $t'$ denote the intra- and inter-cell nearest-neighbor hoppings respectively, $t_3$ represents the next-nearest-neighbor hopping, and $\gamma$ introduces the non-Hermiticity via hopping between sublattices. When $t_3=0$, the Eq. \ref{pt-nh} refers to the Hamiltonian without next-nearest-neighbor hopping, and the critical points for the system under PBCs are deduced:
\begin{equation}
t= \pm t^{\prime} \pm (\frac{\gamma}{2}),
\label{PBCt}
\end{equation}

   We set $t^{\prime} = 1.0$ and $\gamma = 4/3$, yielding a critical point of the phase transition at $t \approx 1.67$ \cite{yao2018edge}. 
  when $t_3$ $\neq 0$,  the topogical transition evidently shifts backward, due to the effect of the next-nearest-neighbor hopping; though an analytical expression for phase boundary is not available, the  phase  transition  can  be  confirmed  through observing the divergences of $T_d$ and its partial derivatives, as demonstrated in Fig. 2, showing a good agreement with  Ref \cite{yao2018edge}.

~\\

{\it Open Boundary Conditions.-} Then  we study  the  non-Hermitian  Halmitonian  under OBCs, and the bulk spectra of an open system evidently differ from that of a periodic one, revealing a collapse of the conventional bulk-boundary correspondence. Notably, the eigenstate  becomes exponentially localized near the boundaries, thus Bloch factor of the wavefunction lies along a certain trajectory in the complex plane, allowing the  presence of GBZ, displayed in Fig. 3(a).

For non-Hermitian system under OBC, it is necessary to transform $k$ to its complex form, i.e., $k \rightarrow k- i\ln r$. When $t_3=0$, the critical points for the Hamiltonian without next-nearest-neighbor hopping  are deduced.

\begin{align}
	t= \pm \sqrt{ \pm(t^{\prime})^2+(\frac{\gamma}{2})^2}.
	\label{OBCt}
\end{align}

The parameters are set to be $t^{\prime} = 1.0$ and $\gamma=4/3$, and we observe a dramatic divergence around the phase transition $t \approx 1.20$, as shown in Fig. 3(b)-(d). When $t_3$ $\neq 0$, we confirm that the phase boundary locates at $t \approx 1.56$, agreeing well with Ref\cite{yao2018edge}, and suggesting that the method is validated and keeps robust against the variation introduced by the next-nearest-neighbor hopping.

~\\

{\it Higher‑Order Topological Phases in Non‑Hermitian Systems.-}Higher-order topology has induced considerable interest, and it exhibits different topological properties, in contrast to the first-order topology. For example, a d-dimensional second-order topological insulator (SOTI) only hosts topologically protected ($d$-$2$)-dimensional gapless boundary states. In rencet years, the concept of higher-order topological phase has been extended to non-Hermitian systems, where the conventional bulk-boundary correspondence is no longer applicable. In order to verify the validity of TD in characterizing higher-order phase transitions, we employ 2D and 3D second-order topological insulators (SOTIs) in the presence of non-Hermiticity \cite{liu2019second}.

\begin{figure}
	\centering
	\includegraphics[width=0.48\textwidth]{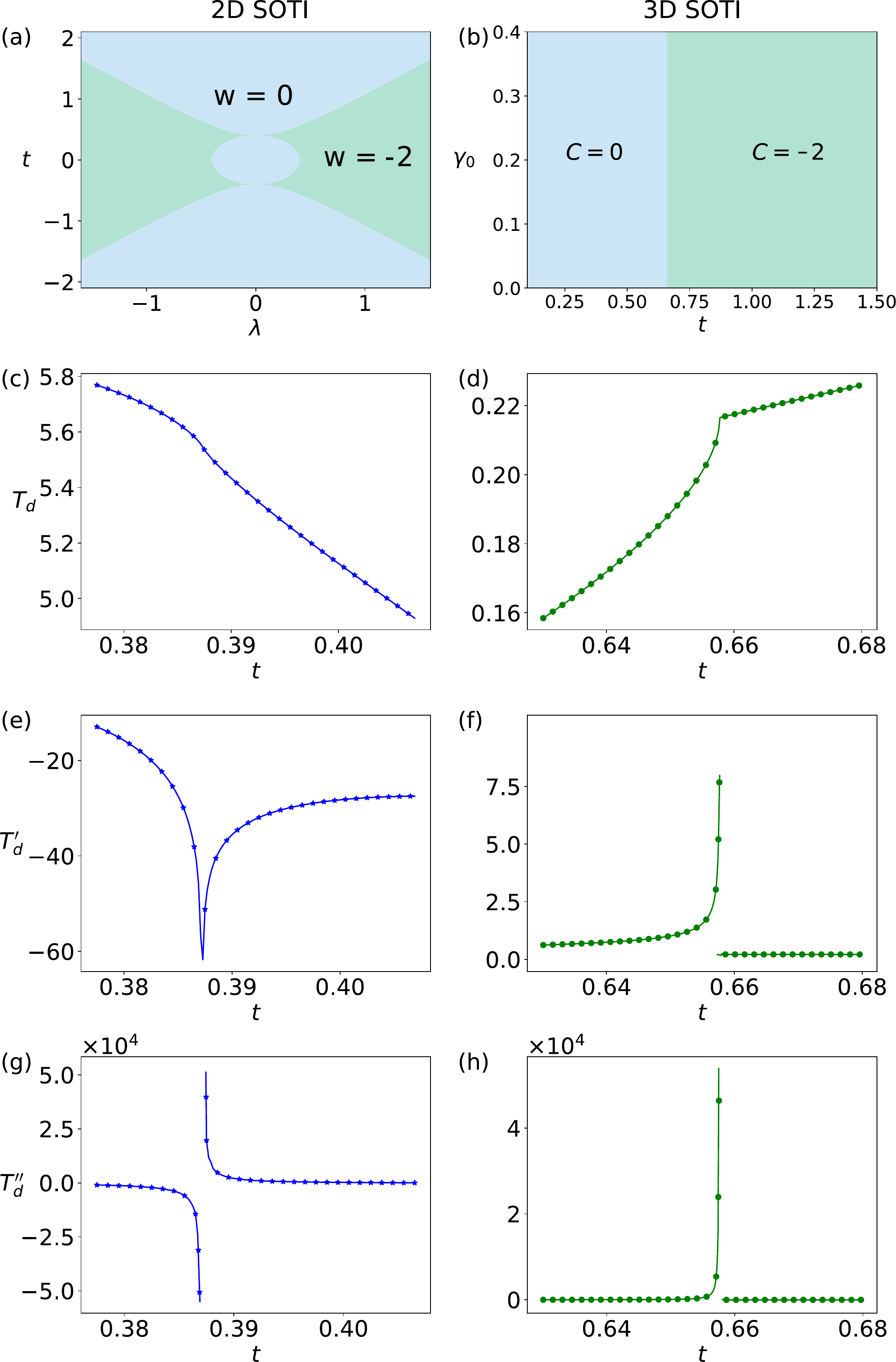}
	\caption{Topological phase diagram, $T_d$ and its partial derivatives of the models in non-Hermitian SOTI; the  left  panel corresponds to 2D SOTI and the right refers to 3D SOTI. (a) Topological phase diagram of the non-Hermitian SOTI in 2D, with the green regions mark the second-order topological phase that hosts corner states and characterized by the non-Bloch winding number (w = -2). (b) Second-order topological-phase diagram in 3D characterized by the non-Bloch Chern number (C = -2).}
	\label{fig4}
\end{figure}

Here, we study a model of the 2D SOTI defined on a square lattice first, where each unit cell includes four orbitals and asymmetric particle hopping within each unit cell is introduced[50]. The Bloch Hamiltonian is expressed as:

\begin{align}
H_{2\mathrm{D}} = \ & \left[t + \lambda \cos(k_x)\right] \tau_x - \left[\lambda \sin(k_x) + i\gamma\right] \tau_y \sigma_z \\
& + \left[t + \lambda \cos(k_y)\right] \tau_y \sigma_y + \left[\lambda \sin(k_y) + i\gamma\right] \tau_y \sigma_x, \nonumber
\end{align}

where $\lambda$,  $t$ represent the intercell hopping amplitude,  intracell hopping amplitude respectively; the Pauli matrices $\tau_i$ and $\sigma_i$ ($i=x,y,z$) stand for the degrees of freedom within a unit cell; $\gamma$ denote the strength of non-Hermiticity. The non-Hermitian system under OBCs considered here does not support first-order topological phases; instead, it  exhibits second-order boundary modes.   The  boundaries of the second-order topological phase are analytically determined by two conditions:  $t^2 = \lambda^2 + \gamma^2$ and $t^2 = \gamma^2 - \lambda^2$. The phase boundary is given by $t = \sqrt{\lambda^2-\gamma^2}$,  and the second-order topological phase emerges between boundaries where the system hosts corner-localized zero-energy modes under OBCs. 
As  shown  in  Fig. 4,  the  left  panel presents the phase diagram, $T_d$ and its partial derivatives of 2D SOTI, the method remains valid and demonstrates a critical point of the phase transitions locating at $t=0.39$, when we set $\lambda=0.4$ and $\gamma=0.1$\cite{liu2019second}.

Then we consider a model of 3D  non-Hermitian SOTI  on a cubic lattice, whose  Hamiltonian is described as:

\begin{align}
H_{3\mathrm{D}} = & \left( m + t \sum_{i = x,y,z} \cos k_i \right) \tau_z + \sum_{i = x,y,z} \left( \Delta_1 \sin k_i + i\gamma_i \right) \sigma_i \tau_x \nonumber \\
& + \Delta_2 (\cos k_x - \cos k_y) \tau_y,
\end{align}

where  $i$ runs over $x$, $y$ and $z$, and $\gamma_x = \gamma_y = \gamma_0$;   $m$ controls the bulk gap;  $t$ represents nearest-neighbor hopping; $\Delta_1$ governs the amplitude of Hermitian spin-orbit-like hopping and $\Delta_2$  modulates a symmetry-breaking term. When the bulk bands are gapped, it does not support gapless surface states.  For non-Hermitian SOTI in 3D, the method is also valid and we can identify the transition point from the divergences of $T_d$ and its derivatives, and the transition point lies at $t = 0.658$, with $\Delta_1 = 1.2,\Delta_2 = 1.1,m = 3.0$, $\gamma_0 = 0.7$ and $\gamma_z = -0.2$, showing a good agreement with Ref \cite{liu2019second}.

~\\
In summary, we have proposed the concept of TD, acquired through the intergration of trace distance over the entire GBZ, allowing the characterization of phase transitions in non-Hermitian topology.  The TD has been employed to calculate the overall difference between wavefunctions accumulated along momentum direction, and observed the divergence of both TD and its partial derivatives to confirm if there occurs a phase transition, different from the conventional topological invariants.  The method based on TD, has been generalized to topological systems, ranging from 1D NH Kitaev model, 1D PT-symmetric SSH model under PBCs or OBCs, to  second-order NH topology in 2D and 3D, where their GBZ are well defined. Such a method characterizes phase boundaries in a general and effective manner, providing a novel insight to explore the non-Hermitian topological physics.

{\it Acknowledgments.-} This work is supported by Xin Jiang Tian Chi Talent Program (5105240150e) and the National Natural Science Foundation of China (61805162).

\bibliography{ref.bib}
\end{document}